\documentclass[twocolumn, switch]{article} % Method A for two-column formatting

\usepackage{preprint}
\usepackage{amsmath, amsthm, amssymb, amsfonts}

\usepackage[numbers,square,sort&compress]{natbib}
\usepackage[utf8]{inputenc}	% allow utf-8 input
\usepackage[T1]{fontenc}	% use 8-bit T1 fonts
\usepackage{xcolor}		% colors for hyperlinks
\usepackage[colorlinks = true,
linkcolor = purple,
urlcolor  = blue,
citecolor = cyan,
anchorcolor = black]{hyperref}	% Color links to references, figures, etc.
\usepackage{booktabs} 		% professional-quality tables
\usepackage{nicefrac}		% compact symbols for 1/2, etc.
\usepackage{microtype}		% microtypography
\usepackage{lineno}		% Line numbers
\usepackage{float}			% Allows for figures within multicol

\usepackage{lipsum}		%  Filler text

\usepackage{newfloat}
\DeclareFloatingEnvironment[name={Supplementary Figure}]{suppfigure}
\usepackage{sidecap}
\sidecaptionvpos{figure}{c}

\usepackage{titlesec}
\titlespacing\section{0pt}{12pt plus 3pt minus 3pt}{1pt plus 1pt minus 1pt}
\titlespacing\subsection{0pt}{10pt plus 3pt minus 3pt}{1pt plus 1pt minus 1pt}
\titlespacing\subsubsection{0pt}{8pt plus 3pt minus 3pt}{1pt plus 1pt minus 1pt}

\usepackage{graphicx}% Include figure files
\usepackage{dcolumn}% Align table columns on decimal point
\usepackage{bm}% bold math
\usepackage{mathptmx}
\usepackage{etoolbox}
\usepackage{enumerate}
\usepackage{caption}
\usepackage{comment} % adicione para comentario
\usepackage{braket}
\usepackage{float}
\DeclareMathOperator{\sinc}{sinc}
\usepackage{mathdots}

\usepackage{stackengine}
\newcommand\widebar[1]{\stackrel{\rule{0.5em}{0.8pt}}{#1}}

\usepackage{tikz}

\definecolor{lime}{HTML}{A6CE39}
\DeclareRobustCommand{\orcidicon}{
	\begin{tikzpicture}
		\draw[lime, fill=lime] (0,0)
		circle [radius=0.16]
		node[white] {{\fontfamily{qag}\selectfont \tiny ID}};
		\draw[white, fill=white] (-0.0625,0.095)
		circle [radius=0.007];
	\end{tikzpicture}
	\hspace{-2mm}
}
\foreach \x in {A, ..., Z}{\expandafter\xdef\csname orcid\x\endcsname{\noexpand\href{https://orcid.org/\csname orcidauthor\x\endcsname}
		{\noexpand\orcidicon}}
}
\title{Spectroscopy of collective modes in a Bose-Einstein condensate: From single to double excitation periods}

\usepackage{authblk}

\author[1\thanks{\tt{lemachado@usp.br}}]{L. A. Machado\orcidA{}}
\author[1]{L. Madeira\orcidB{}}
\author[1]{M. A. Caracanhas\orcidC{}}
\author[1,2]{V. S. Bagnato\orcidD{}}

\affil[1]{Instituto de Física de São Carlos, Universidade de São Paulo, CP 369, 13560-970 São Carlos, Brazil}
\affil[2]{Department of Biomedical Engineering, Texas A\&M University, College Station, Texas 77843, USA}
\begin{document}

\twocolumn[ % Method A for two-column formatting
  \begin{@twocolumnfalse} % Method A for two-column formatting

\maketitle

\begin{abstract}
Collective modes are coherent excitations in Bose–Einstein condensates (BECs), and their study provides insight into the macroscopic quantum phenomena that govern these systems. Collective mode frequencies can be used to probe the properties of BECs, such as the trap geometry, the interatomic interactions, and the presence of defects; hence, it is essential to develop methods for high-resolution determination of collective mode frequencies. A standard technique consists of a single pulse of an external oscillatory field, which the authors denote Rabi-like due to the analogy with the field of nuclear magnetic resonances. In this work, the authors propose a method to achieve a better resolution than the Rabi-like protocol, which consists of two oscillating fields separated in time, which the authors call Ramsey-like. The authors focus on BECs in harmonic traps, considering mainly the quadrupole and breathing modes for different trap anisotropy values and comparing the results using single and double excitation periods. First, the authors employ a variational approach with a Thomas–Fermi ansatz, which gives rise to dynamical equations that are numerically solved. Then, the authors modeled the problem as a three-level system, offering a concise and alternative description of the same dynamics, enabling the idea of coherent control of these collective modes. Both approaches show that the Ramsey-like protocol provides a better resolution than the Rabi-like. This offers the possibility of measuring collective mode frequencies with higher precision, which is important in experiments where they are used as indirect measurements of properties that are difficult to probe. 
\end{abstract}
%\keywords{First keyword \and Second keyword \and More} % (optional)
\vspace{0.35cm}

  \end{@twocolumnfalse} % Method A for two-column formatting
] % Method A for two-column formatting

%\begin{multicols}{2} % Method B for two-column formatting (doesn't play well with line numbers), comment out if using method A

%%%%%%%%%%%%%%%  Main text   %%%%%%%%%%%%%%%
% \linenumbers
\section{\label{sec:intro} Introduction}

A hallmark of Bose-Einstein condensation is the emergence of a superfluid phase~\cite{griffin1987effect}. This coherent phenomenon has interesting implications, such as the appearance of collective oscillations in trapped superfluids~\cite{dalfovo1999theory, mewes1996collective}. These collective oscillations depend on the properties of the system, such as the trap geometry, atom-atom interactions~\cite{PhysRevA.81.053627}, and the presence of defects~\cite{tomishiyo2024superfluid}. Improving the resolution of measurements of collective mode frequencies is essential to enhancing and designing applications where they can be employed as sensors.

\begin{figure}[!htb]
	\centering
	\includegraphics[width=240pt]{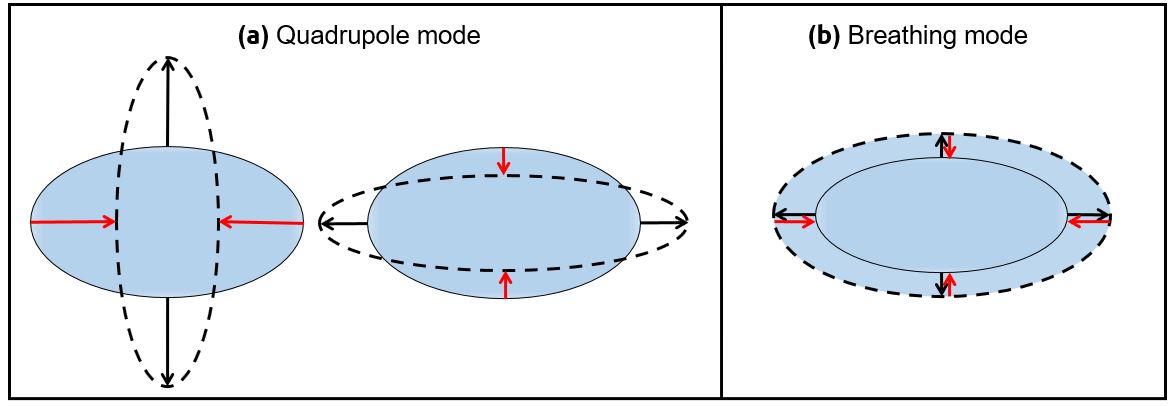}
	\caption{Illustrative representation of $(a)$ the quadrupole mode, exhibiting an out-of-phase oscillation with one axis expanding (black arrow) and the other contracting (red arrow), and $(b)$ the breathing mode, where both axes of the BEC can either expand (black arrow) or contract (red arrow) in phase.}
	\label{fig:colletivemodes}
\end{figure}

Among all the collective modes of a harmonically trapped BEC,
we focus on the quadrupole and breathing modes, as they are among the lowest-energy excitations and are readily excited by modulations of the trapping potential. Furthermore, their discrete spectrum allows for a description in terms of well defined energy levels, which we explore in this work. The quadrupole and breathing modes are characterized by out-of phase and in-phase radii oscillations, respectively, as illustrated in Fig.~\ref{fig:colletivemodes}. 

%%%%%%%%%%%%%%%%%%%%%%%%%%%%%%%%%%%%%%%%%%%%%%%%%%%%%%%%%%%%%%%%%%%%%%%%%%
%%%%%%%%%%%%%%%%%%%%%%%%%%%%%%%%%%%%%%%%%%%%%%%%%%%%%%%%%%%%%%%%%%%%%%%%%%

The main challenge in designing excitation protocols for these collective modes is to produce perturbations that excite predominantly the desired mode while the other modes are suppressed. Commonly used approaches involve time-periodic modulations of the trapping potential~\cite{jin1996collective,mewes1996collective,PhysRevA.56.4855,garcia1999extended} and $s$-wave scattering length modulations using Feshbach resonances~\cite{vidanovic2011nonlinear,PhysRevA.81.053627}. Another approach to excite collective modes involves applying an oscillating magnetic field gradient that overlaps with the trap field~\cite{fritsch2015nonlinear}.

Two well-known excitation protocols in atomic physics are the Rabi~\cite{rabi1938new} and Ramsey techniques~\cite{ramsey1950molecular,ramsey1990experiments,vanier2005atomic} applied in the context of nuclear magnetic resonance to manipulate quantum states through external oscillating fields. The latter is vastly used in high-resolution spectroscopy of atoms~\cite{zanon2018composite} and interferometry~\cite{borde1984optical}. The Rabi technique applies an external field to promote spin transitions, whether electron or nuclear spins. If the driving frequency matches the frequency associated with the energy difference between two quantum states, the population oscillates sinusoidally between them, a phenomenon known as Rabi oscillations~\cite{foot2004atomic}. Therefore, the Rabi technique provides a straightforward way to determine resonance frequencies and control populations between states but has limitations in terms of precision. On the contrary, the Ramsey technique, also known as Ramsey’s method of separated oscillatory fields, improves upon the Rabi method by using two excitation periods separated by a free evolution period rather than a continuous oscillating field. A free evolution period allows the system to acquire a relative phase between the two superposition states, resulting in interference fringes after the second excitation period. These interference fringes, known as Ramsey fringes, narrow the resonance peak, enhancing the precision of frequency measurements compared to the Rabi technique. Moreover, the shorter interaction time with the oscillating field minimizes relaxation and dephasing effects, leading to more accurate and stable measurements, which are crucial in quantum coherence experiments. The question of if this inherent advantage of the Ramsey protocol over the Rabi method is also present in other physical systems arises, which we investigated in this work in the context of BEC collective modes.

In this work, we considered an oscillating potential that overlaps with the trap potential to produce controlled excitations of the collective modes. We employed the concept of excitation periods in analogy with the Rabi and Ramsey techniques, but now, using a many-body system as the target. Using a variational approach, we consistently found that the Ramsey-like excitations provide a better resolution than the Rabi-like method. Moreover, we show a regime where the complex dynamics of the many-body system can be described by a simple three-level model, which makes a straightforward connection to the idea of coherent control of the collective modes.

This work is organized as follows. In Sec.~\ref{sec:collective_modes}, we briefly introduce the collective modes of interest, namely the breathing and quadrupolar modes. Section~\ref{sec:excitation_protocol} contains the details of the Rabi- and Ramsey-like excitation protocols. In Sec.~\ref{sec:variational}, we introduce a variational approach and the numerical solutions to the resulting differential equations. Section~\ref{sec:three_level} deals with the three-level model and its dynamics, which is employed under the light of coherent population control in Sec.~\ref{sec:coherent_control}. Finally, we present our conclusions in Sec.~\ref{sec:conclusions}.

\section{Collective Modes}
\label{sec:collective_modes}

In the mean-field approach, the dynamics of a weakly interacting condensate at zero temperature can be described by the time-dependent Gross–Pitaevskii equation (GPE),
\begin{eqnarray}
	i\hbar \frac{\partial\psi}{\partial t} = \left(-\frac{\hbar^2}{2m}\nabla^2  + V(\mathbf{r},t) + g|\psi(\mathbf{r},t)|^2\right)\psi,
	\label{eq:TDGPE}
\end{eqnarray}
where $g = 4\pi \hbar^2 a_s/m$ with $m$ being the atomic mass and $a_s$ the $s$-wave scattering length. In this work, we considered an axially symmetric harmonic oscillator trapping potential given by
\begin{equation}
	\label{eq:trap}
	V_{\rm trap}(r) = \frac{m \omega_r^2}{2} (x^2 + y^2 + \lambda^2 z^2),
\end{equation}
where $\lambda = \omega_z/\omega_r$ is the anisotropy factor responsible for the shape of the BEC.

Considering the regime of small deviations around the equilibrium density, where the system preserves its coherence, we can assume a linear regime and use the hydrodynamical formalism~\cite{pethick2008bose} to obtain the frequencies of the collective modes. This formalism gives us an alternative description of the dynamics of a BEC based on the dynamics of a quantum fluid. When the typical spatial scales of the system are greater than the wavelength presented by the wave function, then we say the system is in the hydrodynamical regime. In this scenario, by writing the wave function as $\psi(\mathbf{r},t) = \sqrt{n(\mathbf{r},t)}\exp{\left( iS(\mathbf{r},t)\right)}$, where $n(\mathbf{r},t)$ and $S(\mathbf{r},t)$ are, respectively, the density and phase of the BEC, we can express the BEC dynamics in terms of $n(\mathbf{r},t)$ and the superfluid velocity field $\mathbf{v}(\mathbf{r},t)=(\hbar/m)\boldsymbol{\nabla}S(\mathbf{r},t)$. Considering linear perturbations on the density profile around the equilibrium density $n_{eq}(\mathbf{r},t)$ as $n(\mathbf{r},t) = n_{eq}(\mathbf{r},t) + \delta n(\mathbf{r},t)$, where $\delta n(\mathbf{r},t) \ll n_{eq}(\mathbf{r},t)$, and looking for solutions of the form $\delta n (\mathbf{r},t) = \delta n(\mathbf{r})\exp{\left( i\omega t\right)}$ assuming a linear regime, the frequencies of these collective modes are obtained via~\cite{pethick2008bose}
\begin{equation}
\label{eq:EDPfreq}
\omega^2 \delta n = \omega_r^2 \left(r \frac{\partial}{\partial r} + \lambda^2z\frac{\partial}{\partial z} \right) \delta n -  \frac{\omega_r^2}{2}\left( R_r^2 - r^2 - \lambda^2z^2\right)\nabla^2\delta n,
\end{equation}
where $R_r^2$ is the Thomas-Fermi radius along the radial direction.
Among all the possible solutions of Eq.~(\ref{eq:EDPfreq}) given in terms of spherical harmonics $Y_{\ell m}(\theta,\phi)$, we focus on the coupled breathing ($\ell=0,m=0$) and quadrupole ($\ell=2,m=0$) modes. In this regime, the frequency of these low-lying modes is given by~\cite{pethick2008bose}
\begin{equation}
	\omega_{B,Q} = \omega_r\left(2 + \frac{3}{2}\lambda^2 \pm \frac{1}{2}\sqrt{16 - 16\lambda^2 + 9\lambda^4} \right)^{1/2},
	\label{eq:wq}
\end{equation}
where the plus sign is related to the frequency of the breathing mode $\omega_B$, while the minus sign denotes the frequency of the quadrupolar mode $\omega_Q$.

\section{Excitation protocol}
\label{sec:excitation_protocol}

The potential term $V(\mathbf{r},t)$ in Eq.~(\ref{eq:TDGPE}) is the sum of the trapping potential, Eq.~(\ref{eq:trap}), and an external potential responsible for exciting the quadrupole and the breathing modes~\cite{Yukalov2023}. We considered a perturbative potential given by
\begin{equation}
	\label{eq:shapesignal}
	V_{\rm osc}(z,t) =  A f(t)\xi(t) \lambda^2 z^2,
\end{equation}
where $\xi(t)$ represents a modulating signal, and the oscillating function is chosen to be
\begin{equation}
	f(t) = 1 - \cos (\omega_{\rm exc}t),
\end{equation}
introducing the driving frequency $\omega_{\rm exc}$. The constant $A$ determines the perturbation strength and has units of energy per length squared. Excitation potentials such as this one can be created through the composition of magnetic fields, and this particular form is inspired by the experimental setup in Refs.~\cite{henn2008bose,henn2009observation,garcia2022universal}. Although only experiments with the Rabi-like procedure have been reported, we hope this work motivates the usage of the Ramsey technique, which is a straightforward modification of the existing excitation protocol.

Considering the contribution of both the trapping and excitation potentials, we have
\begin{equation}
	V(r,z,t) = \frac{m\omega_r^2}{2}\left[ r^2 + \lambda^2\left(1 + \frac{2A f(t)\xi(t)}{m\omega_r^2}\right)z^2 \right],
\end{equation}
where $r^2 = x^2 + y^2$. This potential describes a time-dependent modulation along the $z$-direction corresponding to either an expansion or a squeezing along this direction, depending on the value of the term inside the parenthesis. 

\begin{figure*}[!htb]
	\includegraphics[width=\textwidth]{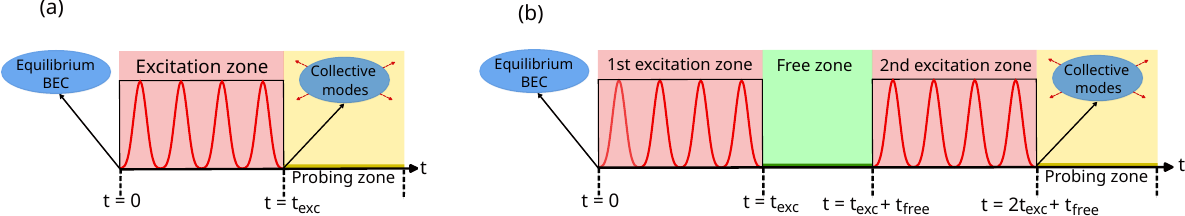}
	\caption{$(a)$ Temporal diagram illustrating the Rabi protocol to excite a BEC. We start with an equilibrium BEC at $t=0$. Subsequently, a single pulse with a duration of $t_{\rm exc}$ is applied. The probing zone is when the excitation protocol ceases, allowing observation of the BEC. $(b)$ Temporal diagram illustrating the Ramsey protocol to excite a BEC. Unlike the Rabi protocol, just after the first excitation zone, the system evolves freely during a time $t_{\rm free}$ until the second excitation zone, also with duration $t_{\rm exc}$, starts at $t=t_{\rm free} + t_{\rm exc}$.}
	\label{fig:protocols}
\end{figure*}

One of the simplest choices for the modulating signal is a single rectangular pulse,
\begin{equation}
	\label{eq:Rabisig}
	\xi(t) =  
	\begin{cases}
		1, & 0 \leq t \leq t_{exc}, \\
		0, & \text{otherwise}, \\
	\end{cases}
\end{equation}
where $t_{\rm exc}$ is the duration of the excitation, as illustrated in Fig.~\ref{fig:protocols}(a). We denote this as the Rabi-like signal, in analogy with the work of Rabi in the field of spin magnetic resonance~\cite{rabi1938new}.

Following the same analogy, we can define a Ramsey signal, drawing a parallel with what Ramsey did to enhance the resolution of the Rabi technique~\cite{ramsey1950molecular}. The Ramsey signal can be written as
\begin{equation}
	\label{eq:Ramseysig}
	\xi(t) =  
	\begin{cases}
		1, & 0 \leq t \leq t_{exc}, \\
		0, & t_{exc} < t \leq t_{exc} + t_{free},\\
		1, & t_{exc} + t_{free} < t \leq 2t_{exc} + t_{free}, \\
		0, & \text{otherwise},
	\end{cases}
\end{equation}
where $t_{\rm free}$ is the interval between the two perturbative pulses when the system evolves freely. The Ramsey protocol involves the separation of the Rabi pulse into two distinct excitation periods, as illustrated in Fig.~\ref{fig:protocols}(b). It is important to note that in the Ramsey excitation scheme, the second pulse is applied at $t_{exc}+t_{free}$. If this interval is not an integer multiple of the driving period $2\pi/\omega_{exc}$, the second pulse may introduce a phase discontinuity relative to the free evolution state. Similarly, the first pulse may also experience a sudden switch-off effect depending on how many oscillations fit within the excitation time $t_{exc}$. These effects are intrinsic to the Ramsey method and play a crucial role in shaping the interference fringes. During the free evolution time $t_{free}$, the populations of collective modes interfere, giving rise to a phase difference, directly influencing the final population distribution, leading to the characteristic Ramsey fringes pattern.

\section{Variational method}
\label{sec:variational}

\subsection{Thomas-Fermi ansatz}

This work assumes a zero-temperature regime corresponding to a superfluid cloud with no thermal component. We start by considering a trial wave function based on the Thomas-Fermi profile~\cite{pethick2008bose,PhysRevA.81.053627},
\begin{flalign}
	\hspace*{-1cm} \nonumber \psi(\mathbf{r},t) = \sqrt{n_0 \left(1 - \frac{x^2}{R_x^2} - \frac{y^2}{R_y^2} - \frac{z^2}{R_z^2} \right)} e^{i\left(\beta_1 x^2 + \beta_2 y^2 + \beta_3 z^2\right) } \nonumber \hspace*{-1cm} \\[0.3cm] \hspace*{-3cm} \times \Theta\left(1 - \frac{x^2}{R_x^2} - \frac{y^2}{R_y^2} - \frac{z^2}{R_z^2} \right), \hspace*{1cm}
	\label{eq:trial}
\end{flalign}
where $R_i$ are the Thomas-Fermi radii, $\Theta$ is the Heaviside step function, the peak density is given by $n_0 = 15N/(8\pi R_xR_yR_z)$, and we have introduced six time-dependent variational parameters: $\{R_i,\beta_i\}$.
Since the phase of the wave function encodes the velocity field of the condensate, it is essential to allow phase dynamics to describe collective oscillations. Our trial wave function in Eq.~(\ref{eq:trial}) extends the Thomas-Fermi ansatz by introducing a phase term that depends on time and space. This choice provides the necessary degrees of freedom for capturing the dynamics of low-lying collective modes~\cite{desaix1991variational}.

Using the trial wave function of Eq.~(\ref{eq:trial}), we applied the variational method~\cite{zooler,ramos2012coupling} to obtain dynamical equations for the BEC radii. The variational method consists in minimizing the Lagrangian $L = \int \mathcal{L}d^3\mathbf{r}$, where 
\begin{equation}
    \mathcal{L} = \frac{i\hbar}{2}\left(\psi^{*}\frac{\partial \psi}{\partial t} - \psi\frac{\partial \psi^{*}}{\partial t}\right) - \varepsilon[\psi]
\end{equation}
is the Lagrangian density with
\begin{equation}
    \label{eq:energyfunc}
    \varepsilon[\psi] = \frac{\hbar^2}{2m}|\nabla \psi|^2 + V(\mathbf{r},t)|\psi|^2 + \frac{g}{2}|\psi|^4
\end{equation}
being the energy functional of the GPE. The gradient term on the right-hand side of Eq.~(\ref{eq:energyfunc}), associated with the kinetic contribution, can be expressed as $|\boldsymbol{\nabla}\psi|^2 = |\boldsymbol{\nabla}\psi_0|^2 + n(\mathbf{r},t)|\boldsymbol{\nabla}S(\mathbf{r},t)|^2$, where we have used the decomposition $\psi(\mathbf{r},t) = \psi_0\exp{\left( iS(\mathbf{r},t)\right)}$. The first term, $|\boldsymbol{\nabla}\psi_0|^2$, accounts for the kinetic energy arising from variations in the wave function amplitude and is associated with the quantum pressure. However, this term is negligible when compared to the phase variations. Consequently, we have neglected $|\boldsymbol{\nabla}\psi_0|^2$ in our calculations. 

The least action principle states that the Lagrangian minimization is achieved by applying the Euler-Lagrange equations with respect to each variation parameter $\{R_i, \beta_i\}$. With this procedure, we end up with the dynamical equations for the BEC radii given by
\begin{gather}
	\label{eq:Rx} \frac{\partial^2\widebar{R}_x}{\partial \tau^2} + \widebar{R}_x = \frac{P_0}{\widebar{R}_x^2 \widebar{R}_y \widebar{R}_z},  \\ 
	\label{eq:Ry} \frac{\partial^2\widebar{R}_y}{\partial \tau^2} + \widebar{R}_y = \frac{P_0}{\widebar{R}_x \widebar{R}_y^2 \widebar{R}_z}, \\ 
	\label{eq:Rz} \frac{\partial^2\widebar{R}_z}{\partial \tau^2} + \lambda^2 \widebar{R}_z = \frac{P_0}{\widebar{R}_x \widebar{R}_y \widebar{R}_z^2}-\lambda^2 \widebar{A} f(\tau)\xi(\tau) \widebar{R}_z,    
\end{gather}
written in a dimensionless form. This is achieved by expressing lengths in terms of the harmonic oscillator length
$\ell_0 = \left( \hbar / m\omega_r\right)^{1/2}$, energies in units of $\hbar \omega_r$, and time in $\omega_r^{-1}$, leading to $\widebar{R}_i = R_i/\ell_0$, $\tau = t\omega_r$, $\widebar{A} = A  \ell_0^2/\hbar\omega_r$, and $P_0 = 15N a_s/\ell_0$. These equations are consistent with Ref.~\cite{castin1996bose}.

This approach gives us Newton-like equations, Eqs.~(\ref{eq:Rx})--(\ref{eq:Rz}), governing the BEC radii with straightforward interpretation. The left-hand side of the equations corresponds to the harmonic oscillator components. The terms proportional to $P_0$ are due to the interatomic interactions, which can either localize or expand the cloud, depending on whether they are attractive or repulsive. The term proportional to $\widebar{A}$, only present in the dynamical equation for the $z$-direction, denotes the role of the external excitation.

Since our approach relies on the Thomas-Fermi approximation, it captures only the overall condensate dynamics and does not resolve excitations with significant spatial structure. This limitation is evident in Eqs.~(\ref{eq:Rx})--(\ref{eq:Rz}), where only the condensate radii appear. Consequently, quantum fluctuations are not considered, and all observed effects arise from linear and nonlinear collective excitations. A more complete description of higher-energy modes would require an extended model incorporating shorter wavelength excitations or a refined variational ansatz that includes additional degrees of freedom to capture finer spatial modulations. However, including these extra degrees of freedom into our ansatz is a nontrivial task, as it requires prior knowledge of the characteristics of these short-wavelength oscillations. An alternative approach to probing higher-energy modes is through the \textit{Bogoliubov–de Gennes} (BdG) method \cite{zhu2016bogoliubov}, which enables the resolution of the entire spectrum of linear excitations. Nevertheless, obtaining analytical solutions to the BdG equations is a hard task. As a result, numerical solutions of the GPE are typically required as a prerequisite for applying the BdG method effectively.

\subsection{Fixed excitation frequency}

We numerically solved Eqs.~(\ref{eq:Rx})--(\ref{eq:Rz}) by employing a fourth-order Runge-Kutta method. Our starting point is a $^{87}$Rb BEC with $N=10^5$ atoms, a $s$-wave scattering length of $a_s = 100a_0$, and a radial frequency of the trap $\omega_r = 2\pi \times 200$ Hz. We obtained the initial conditions by solving the Eqs.~(\ref{eq:Rx})--(\ref{eq:Rz}) neglecting the time derivatives. Since our system has cylindrical symmetry, the time evolution of both $\bar{R}_x$ and $\bar{R}_y$ is the same. Thus, hereafter, they were substituted by $\bar{R}_r$.

\begin{table*}
	\centering
	\caption{Quadrupole and breathing mode frequencies according to the hydrodynamical [Eq.~(\ref{eq:wq})], variational, and three-level models for different values of the anisotropy $\lambda$. For the latter two, the frequencies and their associated uncertainties are obtained from the resonance curves fitted to the functional forms of Eq.~(\ref{eq:fitRabi}) and Eq.~(\ref{eq:fitRamsey}) for the Rabi and Ramsey protocols, respectively.}
	\begin{tabular}{ccccccllllccccc}
		\hline
		& \multicolumn{5}{c}{Quadrupole mode} &  &  &  &  &   \multicolumn{5}{c}{Breathing mode}         \\ \hline
		& Eq.~(\ref{eq:wq})                                 & \multicolumn{2}{c}{Variational}     & \multicolumn{2}{c}{Three-level}     &  &  &  &  &  Eq.~(\ref{eq:wq})                                & \multicolumn{2}{c}{Variational}     & \multicolumn{2}{c}{Three-level}     \\ \hline
		&  & Rabi             & Ramsey           & Rabi             & Ramsey           &  &  &  &  &  & Rabi             & Ramsey           & Rabi             & Ramsey           \\ \hline
		$\lambda$ & $\bar{\omega}_Q$                 & $\bar{\omega}_Q$ & $\bar{\omega}_Q$ & $\bar{\omega}_Q$ & $\bar{\omega}_Q$ &  &  &  &  & $\bar{\omega}_B$                 & $\bar{\omega}_B$ & $\bar{\omega}_B$ & $\bar{\omega}_B$ & $\bar{\omega}_B$ \\ \hline
		$1/9$     & $0.18$                           & $0.2(2)$         & $0.18(6)$        & $0.2(2)$         & $0.18(6)$        &  &  &  &  & $2.00$                           & $2.0(2)$         & $2.00(7)$        & $2.0(2)$         & $2.00(6)$        \\ \hline
		$1/2$     & $0.77$                           & $0.8(2)$         & $0.77(7)$        & $0.8(1)$         & $0.77(6)$        &  &  &  &  & $2.03$                           & $2.1(2)$         & $2.03(7)$        & $2.0(1)$         & $2.03(7)$        \\ \hline
		$1$       & $1.41$                           & $1.5(2)$         & $1.45(6)$        & $1.4(1)$         & $1.41(4)$        &  &  &  &  & $2.24$                           & $2.3(2)$         & $2.24(6)$        & $2.2(2)$         & $2.24(6)$        \\ \hline
		$2$       & $1.76$                           & $1.7(2)$         & $1.74(6)$        & $1.7(2)$         & -----            &  &  &  &  & $3.59$                           & $3.7(2)$         & $3.71(8)$        & $3.6(5)$         & -----            \\ \hline
	\end{tabular}
	\label{tab:tabw}
\end{table*}

\begin{figure*}[!htb]
	\includegraphics[width=\textwidth]{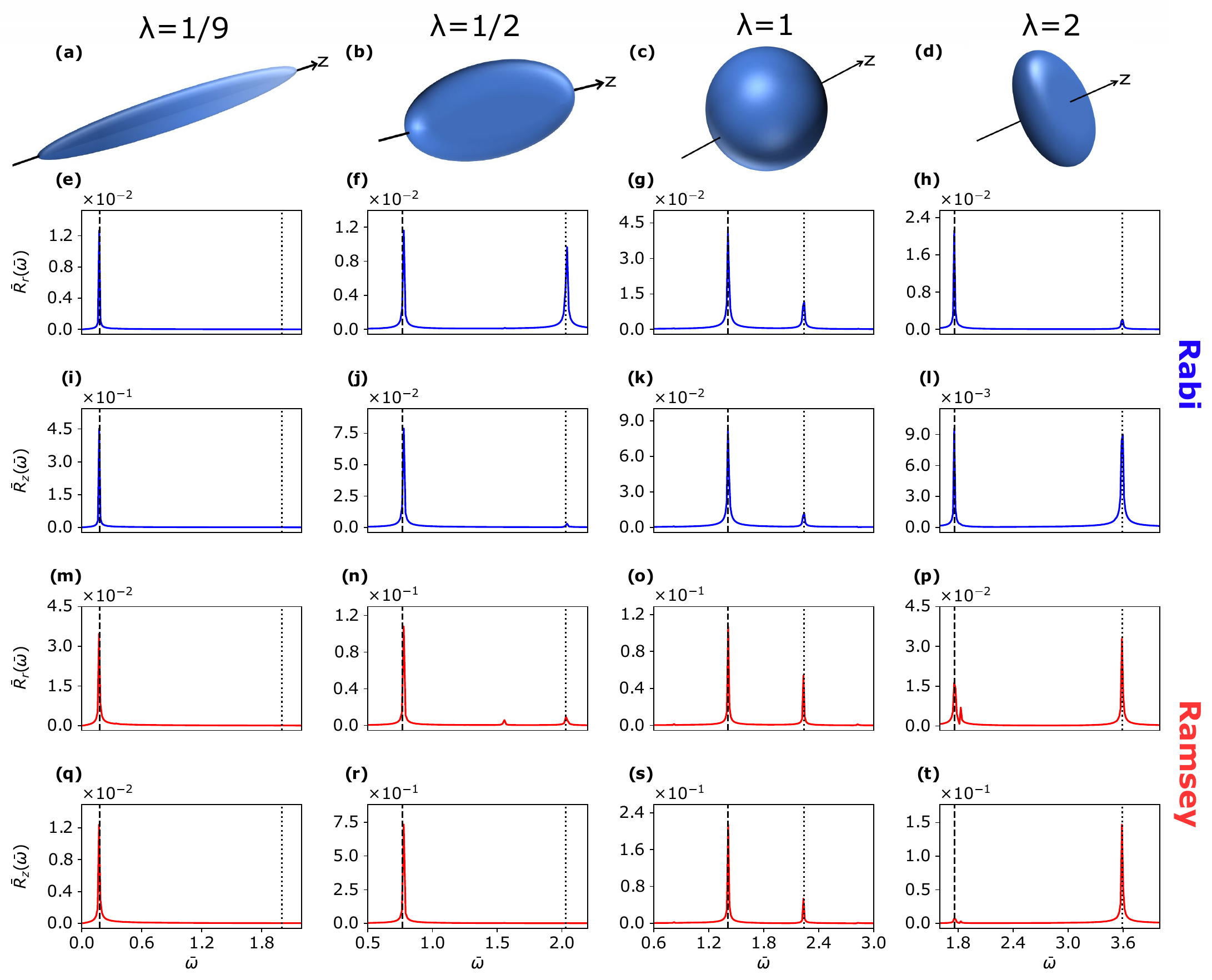}
	\caption{
    $(a-d)$ Illustrations of the equilibrium BEC for different values of the anisotropy $\lambda$, ranging from a prolate ($\lambda < 1$) to an oblate ($\lambda > 1$) geometry.
	$(e-t)$ Fourier transforms of the radii oscillations of the BEC. Two protocols were used to excite the collective modes: the Rabi-like is denoted by blue curves $(e-l)$, while the red curves indicate the Ramsey-like protocol $(m-t)$. We employed the excitation parameters: $\widebar{A} = 0.05$, $\widebar{\omega}_{\rm exc} = 1$, $\tau_{\rm exc} = 32$, and $\tau_{\rm free} = 32$. Each column corresponds to a different trap anisotropy, $\lambda=1/9$, $1/2$, 1, and 2. The dashed and dotted lines correspond to the expected frequency for the quadrupole and breathing modes, respectively, according to Eq.~(\ref{eq:wq}).}
	\label{fig:FFT_osc}
\end{figure*}

We fixed the driving frequency $\widebar{\omega}_{\rm exc} = 1$ for this first investigation, while the other excitation parameters correspond to $\widebar{A} = 0.05$ and $\tau_{\rm exc} = \tau_{\rm free} = 32$. In Tab.~\ref{tab:tabw}, we provide the quadrupole and breathing frequencies according to Eq.~(\ref{eq:wq}) for different values of $\lambda$.

To measure the frequencies of the collective modes, we investigate the evolution of the BEC radii after the perturbation has ceased for a fixed excitation frequency $\widebar{\omega}_{\rm exc} = 1$. Figure~\ref{fig:FFT_osc} shows the Fourier Transform (FT) of the time evolution of the BEC radii after the excitation protocol using both Rabi and Ramsey signals. The frequency resolution in our numerical simulations is limited by the duration of the probing period $\tau_p$, which was set to approximately fourteen times the excitation time, following typical experimental conditions. In our case, the probing period duration was $\tau_p = 468$, leading to a frequency step of approximately $\Delta\bar{\omega} = 0.0134$. Although increasing the probing time could enhance frequency resolution, experimental constraints -such as finite observation windows due to the lifetime of a BEC- limit the duration of such measurements. The peaks depend on the modulation of the external signal, indicating different coupling to the collective modes. Although some cases display multiple peaks, the ones with higher amplitudes are at the expected frequencies for the quadrupole and breathing modes.

Figure~\ref{fig:FFT_osc} shows that the perturbation excites the quadrupole mode more than the breathing mode for $\lambda \leqslant 1$. In the case of $\lambda = 1/9$ [Figs.~\ref{fig:FFT_osc}(e), (i), (m), and (q)], the breathing mode is barely excited. As we decrease the anisotropy of the trap, the amplitude of the breathing mode becomes appreciable, although the quadrupole mode is still more relevant for $\lambda \leqslant 1$. This can be attributed to two main factors:  the excitation was applied along the $z$-direction, favoring an out-of-phase oscillation; the quadrupole mode is more easily excited due to the lower confinement in the $z$-direction. An interesting scenario is when there is an inversion of the confining strength. For $\lambda = 2$ [Figs.~\ref{fig:FFT_osc}(h), (l), (p), and (t)], the radial direction becomes more confining than the $z$-direction, and the amplitude of the breathing mode can be of the order of the quadrupole mode and even larger in some cases.

\begin{figure}[!htb]
	\centering
	\includegraphics[width=250pt]{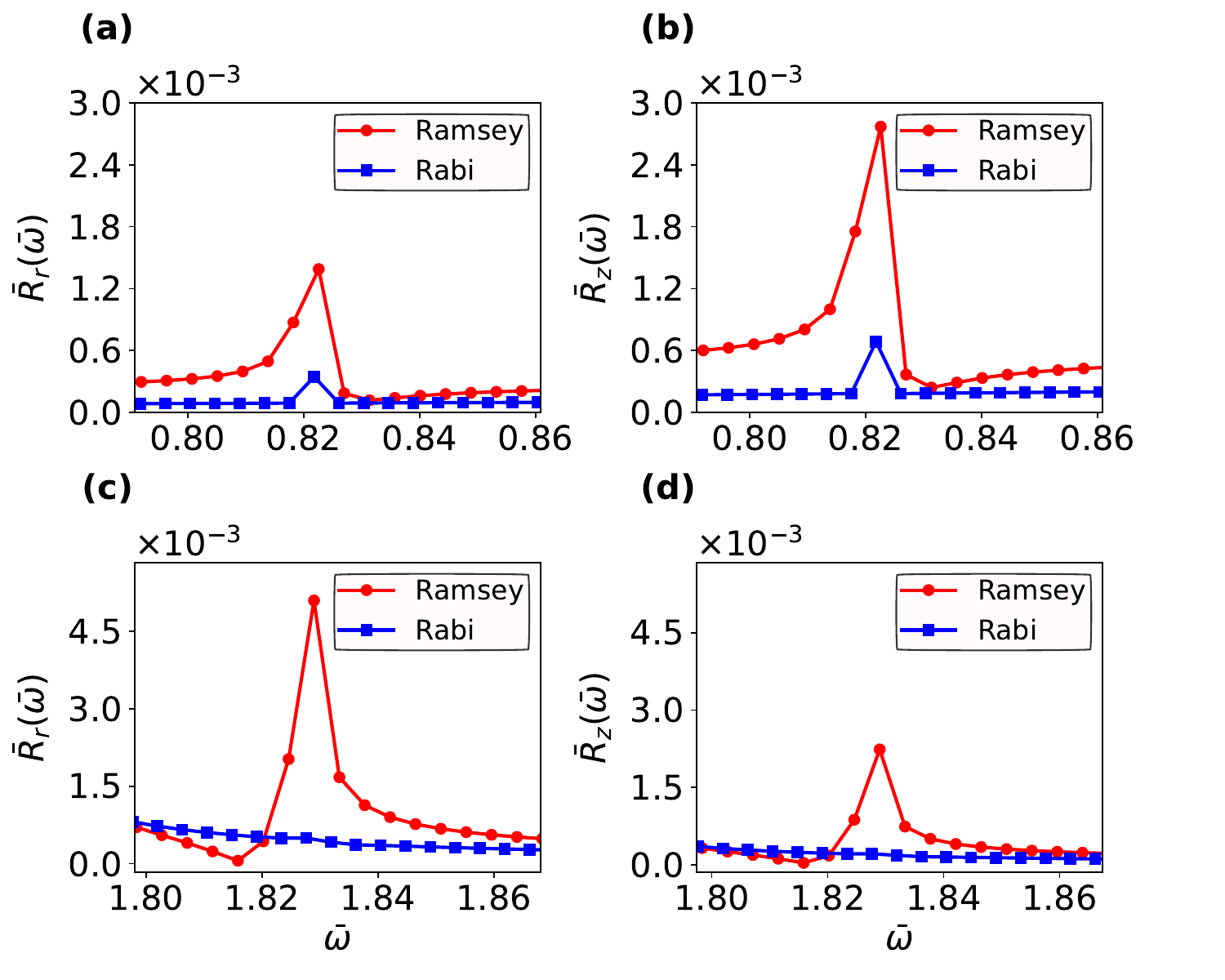}
	\caption{Fourier transforms of the radii oscillations of the BEC using the Rabi and the Ramsey protocols for $\lambda = 1$ $(a-b)$ and $\lambda=2$ $(c-d)$, evidencing the contribution of a mode with a frequency given by the difference between the breathing and quadrupole mode frequencies, $\widebar{\omega}' = \widebar{\omega}_B - \widebar{\omega}_Q$. This shows the Ramsey method provides increased sensitivity}, enabling the observation of modes more clearly than its Rabi counterpart.
 
	\label{fig:inter_peaks}
\end{figure}

There are peaks at positions other than the quadrupole and breathing frequencies. Indeed, the emergence of nonlinear mode-mixing in trapped BECs has already been discussed in references \cite{PhysRevA.56.4855,vidanovic2011nonlinear}. Nevertheless, its noteworthy the additional peak with frequency $\widebar{\omega}' = \widebar{\omega}_B - \widebar{\omega}_Q$, i.e., an intermediate state between the breathing and quadrupole modes transition. For $\lambda=2$, where $\widebar{\omega}'= 1.83$ is close to the quadrupole frequency %(generating beating effects in the radii oscillations)
, it can be seen in Fig.~\ref{fig:FFT_osc}(p) and, less pronounced, in (t). In the case of $\lambda = 1$, the peak at the frequency $\widebar{\omega}' = 0.82$ has a much smaller magnitude. In Fig.~\ref{fig:inter_peaks}, we zoom in on relevant regions for better visualization and compare the Rabi and Ramsey protocols. Since we aim to highlight a relatively narrow interval, we increased the probing time $\tau_p$ in our numerical calculation so that the frequency step is approximately $\Delta\bar{\omega} = 0.0043$. This example illustrates the advantage of the Ramsey-like protocol over the Rabi-like since these peaks appear more clearly using the double excitation periods protocol, indicating that this signal also offers greater sensitivity to probe this additional peak.

To complete our understanding of the types of peaks present in Fig.~\ref{fig:FFT_osc}, we should also discuss the additional peak present in Fig.~\ref{fig:FFT_osc}(n) at $\bar{\omega} = 1.54$, which is twice the frequency of the quadrupole mode for $\lambda=1/2$. Its manifestation occurs due to the non-linear contribution of the hydrodynamical equations~\cite{guo2004second,hechenblaikner2000observation}, a process similar to what happens in non-linear optics when there is the second harmonic generation~\cite{campagnola1999high}, resulting in multiples of the collective mode frequency.

\subsection{Resonance curves}

As previously discussed, the Ramsey method offers a significant advantage over the Rabi method in achieving higher precision in measuring system resonances. In this section, we aim to construct the resonance curves of the collective modes using both the Rabi and Ramsey excitation protocols, subsequently comparing the accuracy provided by each approach.

To construct the resonance curves, we have to repeat the procedure of Fig.~\ref{fig:FFT_osc}, which was obtained with $\widebar{\omega}_{\rm exc}=1$, for other values of the excitation frequency. Then, we select the maximum amplitude observed in the FT for a given mode (quadrupole or breathing), radius ($R_r$ or $R_z$), anisotropy factor, signal, and excitation frequency. This becomes one point in the resonance curve; the others are obtained by varying the excitation frequency. We adopted the same excitation parameters as in the case of Fig.~\ref{fig:FFT_osc}, i.e., we set $\widebar{A} = 0.05$ and $\tau_{\rm exc} = \tau_{\rm free} = 32$, while we varied the excitation frequency $\widebar{\omega}_{\rm exc}$ in the range from $0.1$ to $4.2$. These parameters were chosen to obtain a clear interference pattern in the resonance curve, to keep the system in the so-called coherent regime (i.e., where the excited modes are mainly the desired collective modes), and to be compatible with typical experiments~\cite{henn2008bose,garcia2022universal}.

\begin{figure*}[!htb]
	\includegraphics[width=\textwidth]{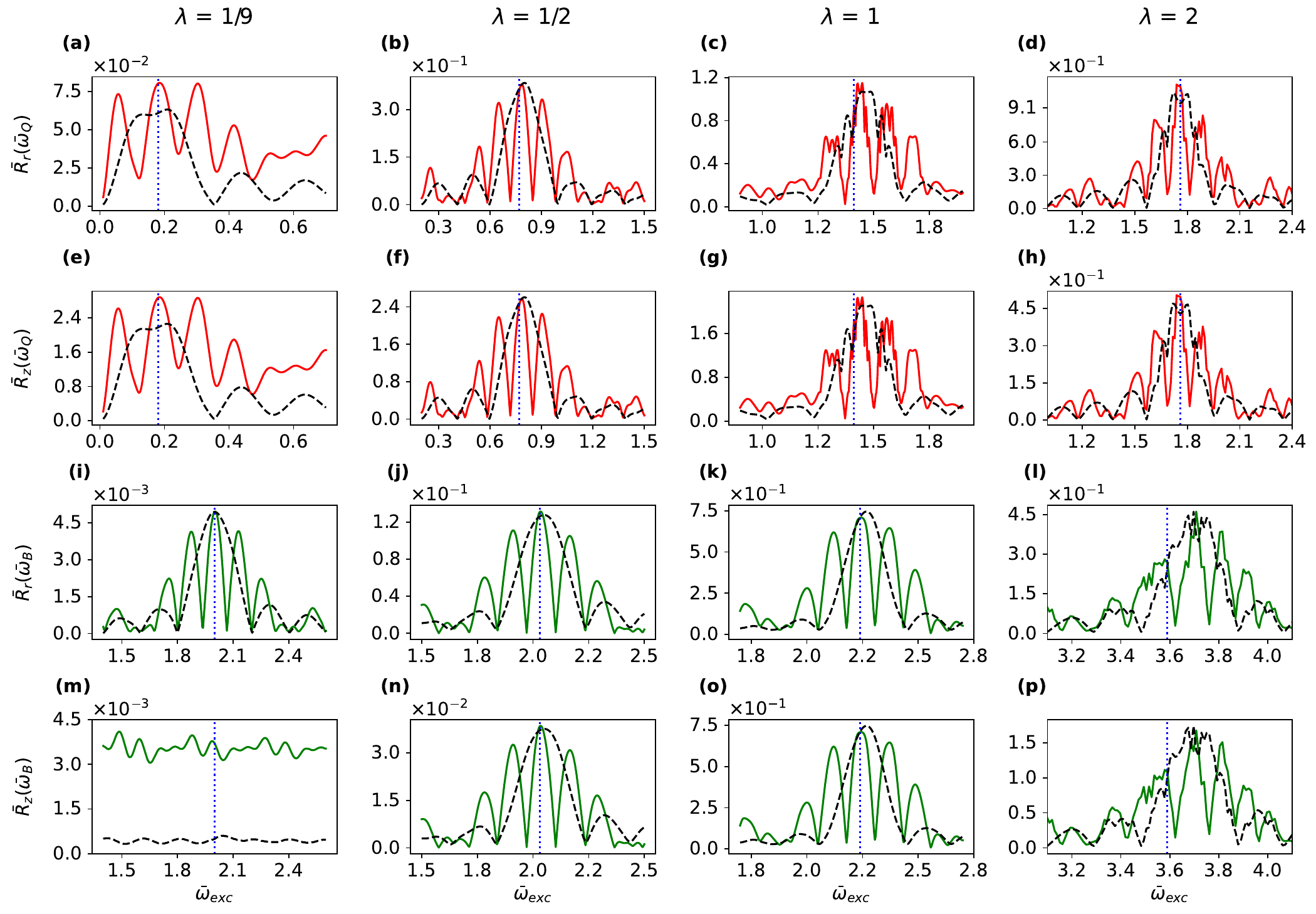}
	\caption{
 Resonance curves for the $(a-h)$ quadrupole and $(i-p)$ breathing modes. Dashed black curves denote the Rabi excitation protocol, while the Ramsey technique is represented by colored curves (red for the quadrupole and green for the breathing modes). The used excitation parameters were $\widebar{A} = 0.05$ and $\tau_{\rm exc} = \tau_{\rm free} = 32$. The vertical dotted lines are at the expected resonance frequencies according to Eq.~(\ref{eq:wq}). Each column corresponds to a different trap anisotropy, $\lambda=1/9$, $1/2$, 1, and 2.
 }
	\label{fig:ress_vari}
\end{figure*}

Figure~\ref{fig:ress_vari} summarizes our findings. The most notable feature is the appearance of Ramsey fringes when the protocol with the two excitation periods is applied, in analogy with Ramsey's interferometry method~\cite{ramsey1990experiments}. The consequence is that, both in interferometry and in our case, the Ramsey method yields narrower peaks than the Rabi protocol, providing better accuracy in frequency measurements. This is seen in Fig.~\ref{fig:ress_vari}[(b), (f), (j), and (n)]. 

Next, we need a method for extracting the collective modes frequencies and their uncertainties from the resonance curves, for which we employ again the overarching analogy of this work. For a two-level quantum system subject to a sinusoidal-like external perturbation, the solution for the transition probability is straightforwardly obtained for both Rabi and Ramsey cases~\cite{foot2004atomic}. For the Rabi protocol, the solution has a functional form given by
\begin{equation}
R_{\rm Rabi}(\Delta) = A\left( \frac{\sigma^2}{\Delta^2 + \sigma^2} \right)\sin^2{\left(B\sqrt{\sigma^2+\Delta^2}\right)},
\label{eq:fitRabi}
\end{equation}
which is a Lorentzian modulated by the $\sin^2(B\sqrt{\sigma^2+\Delta^2})$ function. The parameters to be adjusted are $A$, $B$, $\sigma$, and $\Delta = \bar{\omega}_{exc} - \bar{\omega}_{Q,B}$, which is the detuning. We estimate the uncertainty as the full width at half maximum (FWHM) of the curve, which leads to the transcendental equation
\begin{equation}
\label{eq:rabi_trans}
	\sin{\left(B\sqrt{\sigma^2 + \Delta^2}\right)} = \frac{\sqrt{\sigma^2 + \Delta^2}}{\sqrt{2A}\sigma}. 
\end{equation}
The value of $\Delta$ that satisfies Eq.~(\ref{eq:rabi_trans}), which we denote $\Delta_0$, is then used to compute the uncertainty:
\begin{equation}
	\delta \bar{\omega}_{\rm Rabi} = 2\Delta_0. 
\end{equation}
We considered only the frequency range of the main peak to compute the FWHM.

Similarly, the Ramsey functional form is given by~\cite{foot2004atomic}
\begin{equation}
	R_{\rm Ramsey}(\Delta) = A'\sinc^2{\left( B'\Delta\right)}\cos^2{\left( C'\Delta\right)},
	\label{eq:fitRamsey}
\end{equation}
which resembles a Rabi function, but the modulation is given by the factor $\cos^2(C'\Delta)$. The fitting parameters correspond to $A'$, $B'$, and $C'$. The $\cos^2(C'\Delta)$ term generates the interference fringes; thus, the amplitude falls to its half when $C'\Delta=\pi/4$. Hence, we can obtain the width of the Ramsey resonance curve through
\begin{equation}
	\delta \bar{\omega}_{\rm Ramsey} = \frac{\pi}{2C'}. 
\end{equation}
Indeed, the ideal Ramsey experiment predicts $\delta \bar{\omega}_{\rm Ramsey} = \pi/\tau_{\rm free}$, which for our parameters results in $\delta\bar{\omega}_{\rm Ramsey} = 0.098$.

We employed the radial direction, Fig.~\ref{fig:ress_vari}(a-d) and (i-l), together with the functional forms of Eqs.~(\ref{eq:fitRabi}) and (\ref{eq:fitRamsey}), to extract the collective mode frequencies and their uncertainties from the resonance curves. We summarized our results in Tab.~\ref{tab:tabw}, displayed alongside the other methods used in this work. As expected, for all cases, the Ramsey protocol yields smaller uncertainties than the Rabi method while also being closer to the hydrodynamical result of Eq.~(\ref{eq:wq}).

The $\lambda=1/2$ case [Figs.~\ref{fig:ress_vari}(b), (f), (j), and (n)] is illustrative of differences between the two excitation protocols. The central peak of the interference pattern for $\lambda=1/2$ shows a slight shift between the central peak of Rabi and Ramsey curves, with the latter being closer to the prediction of Eq.~(\ref{eq:wq}).

Although the resonance curves present a much more precise alternative for determining the frequencies of the collective modes than approaches employing only the dynamics resulting from a single excitation frequency, they cannot resolve modes absent in the FT spectrum. This is the case of the breathing mode in the $z$-direction for $\lambda=1/9$, which is barely manifested in the FT spectrum [Figs.~\ref{fig:inter_peaks}(e), (l), (m), (q)]. The consequence is that its associated resonance curve, Fig.~\ref{fig:ress_vari}(m), lacks an interference pattern. Also, for $\lambda=1/9$, when we take the quadrupole mode into account, the resonance pattern is not well-defined. This is primarily because the quadrupole frequency is relatively small, and the chosen excitation time is insufficient to complete even a single excitation period.

Given that the perturbative term in the dynamical Eq.~(\ref{eq:Rz}) scales with $\lambda^2$, the perturbation strength is amplified for $\lambda \geq 1$. In this scenario, the system starts to deviate from the coherent regime as can be seen in the case of the quadrupole mode for $\lambda=1$ [Figs.~\ref{fig:ress_vari}(c) and (g)] and both modes for $\lambda=2$ [Figs.~\ref{fig:ress_vari}(d), (h), (l), and (p)]. Once excitations are amplified, higher energy modes contribute to a nonlinear behavior, leading to large phase fluctuations of the wave function. In these cases, there is an evident shift between the frequencies predicted by Eq.~(\ref{eq:wq}), vertical blue dashed lines in Fig.~\ref{fig:ress_vari}, and the ones extracted from the resonance curves, as shown in Tab.~\ref{tab:tabw}.

Frequency shifts in collective modes have been observed in systems where thermal effects and quantum fluctuations play a significant role~\cite{marago2001temperature, zhou2008frequency,jin1997temperature}. However, since we assume in this work $T=0$ and our equations describe a classical nonlinear field derived under the Thomas-Fermi approximation, both thermal and quantum fluctuations are absent. Yet, we observe frequency shifts in our simulations, which arise due to nonlinear interactions between collective modes. When external perturbations are applied on a spatial scale comparable to the size of the condensate, they primarily excite low-lying collective modes with wavelengths on the order of the BEC radii. However, due to the nonlinearity of the GPE, energy can be transferred between these modes, leading to mode coupling and shifts in their frequencies. For sufficiently strong excitations, energy redistribution extends beyond the lowest modes, exciting additional nonlinear collective modes. In this regime, purely linear collective mode dynamics assumption becomes less accurate as the interactions lead to spectral broadening and frequency shifts. This energy transfer process can diminish the population of the initially excited modes, modifying their spectral response.

\section{Three-level model}
\label{sec:three_level}

\begin{figure}[!htb]
	\centering
	\includegraphics[scale = 0.25]{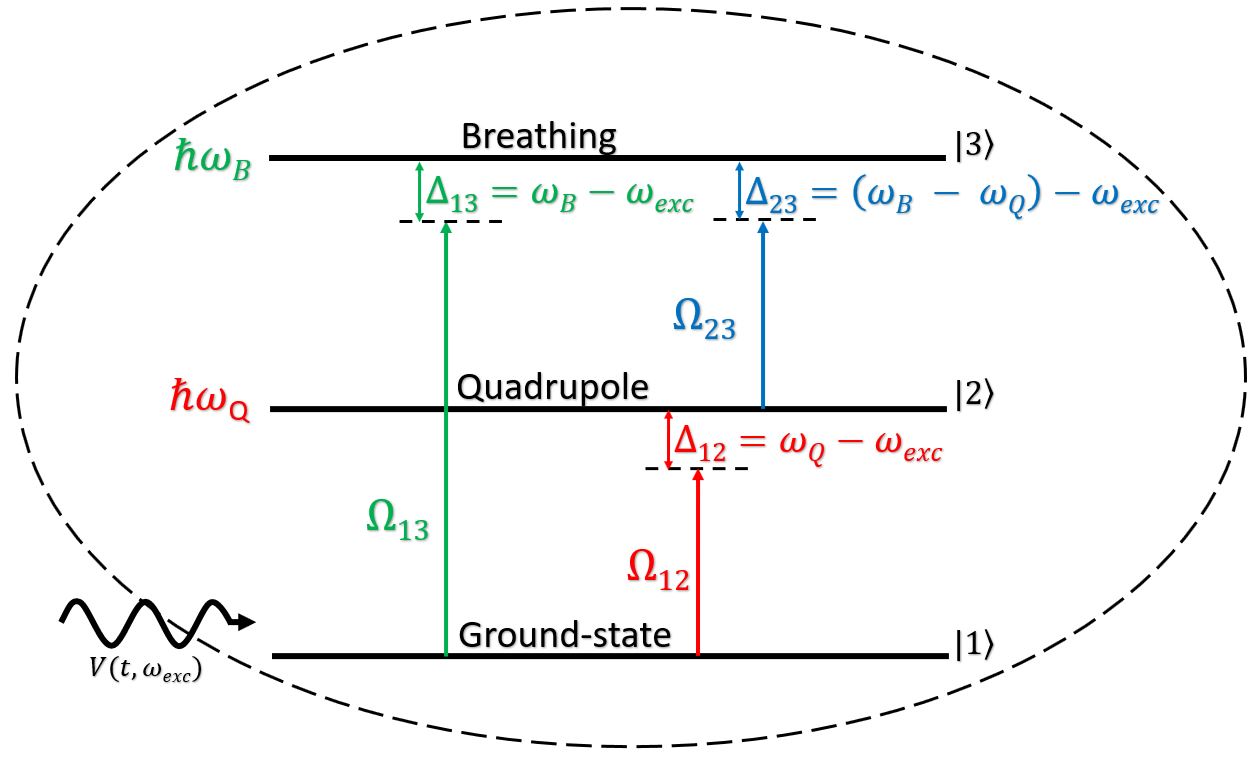}
	\caption{Three-level system under an external perturbation. The upper level is related to the breathing mode, the intermediate level is related to the quadrupole mode, and the lower level stands for the BEC at rest.}
	\label{fig:3lvlscheme}
\end{figure}

Motivated by the results obtained from the numerical solutions of the dynamical equations and the fact that we are considering a regime where the system preserves a high degree of coherence, we propose an alternative description of the dynamics of the collective modes. The resonance curves exhibit narrow widths, supporting the idea of a few discrete quantum states rather than considering a continuous band. In this sense, we propose to model the system as a three-level quantum system, as illustrated in Fig.~\ref{fig:3lvlscheme}, in analogy with a three-level atom subject to an external perturbation. The lowest level, labeled as $\ket{1}$, corresponds to the ground-state BEC, the middle to the quadrupole mode ($\ket{2}$), and the upper level to the breathing mode ($\ket{3}$).

Using the canonical quantization and working in the interaction picture, the perturbative potential in Eq.~(\ref{eq:shapesignal}) is now considered an operator written as
\begin{equation}
	\label{eq:VoscIntPic}
	\hat{V}_{\rm osc}(t) = f(t)\sum_{i,j}\hbar\Omega_{ij}e^{-i(\omega_j - \omega_i)t}\ket{i}\bra{j},
\end{equation}
where we defined Rabi-like frequencies as
\begin{equation}
	\label{eq:defRabif}
	\Omega_{ij}(A,\lambda) = \frac{\lambda^2 A}{\hbar}\bra{i}\hat{z}^2\ket{j}.
\end{equation}
Hence, to determine the Rabi-like frequencies of this system, we have to compute the matrix elements of the operator $\hat{z}^2$. The diagonal matrix elements of this operator are non-zero, in contrast with dipole-like operators~\cite{sakurai2020modern}. Despite this, these elements remain inactive in the dynamics as they do not contribute to the coupling between the modes. They only contribute as a global phase, which does not change the quantum state.

The inner product $\bra{i}\hat{z}^2\ket{j}$ carries significant meaning regarding the coupling of collective modes in the system. Its value indicates which transitions are more likely to occur. Since changes in the geometry of the confining potential alter the coupling of these modes, it is reasonable to expect that this inner product will depend on $\lambda$. Consequently, Eq.~(\ref{eq:defRabif}) will exhibit a dependence on $\lambda$ different from that $\lambda^2$ initially suggested.

This scenario gives rise to three possible transitions $\ket{1} \rightarrow \ket{2}$, $\ket{1} \rightarrow \ket{3}$, and $\ket{2} \rightarrow \ket{3}$, where two of them contribute to the population of the breathing mode. Consequently, one can reasonably anticipate the emergence of two main peaks in the resonance curve of the breathing mode: one positioned at $\omega_B$ and another at $\omega' = \omega_B - \omega_Q$. Each of the three transitions has its own Rabi frequency, reflecting the coupling strength with the external excitation, where higher values of Rabi frequency suggest preferential population transfer to the associated state.

A convenient way to describe the dynamics of the system is through the density operator formalism. We can compute the dynamics of this quantum closed system by evolving the density operator $\hat{\rho}(t) \equiv \ket{\Psi(t)}\bra{\Psi(t)}$ with the master equation~\cite{breuer2002theory,nielsen2010quantum}
\begin{equation}
	\frac{d\hat{\rho}(t)}{dt} = \frac{i}{\hbar}[\hat{\rho}(t),\hat{V}_{\rm osc}(t)].
	\label{eq:masterEq}
\end{equation}
Since the potential $\hat{V}_{\rm osc}(t)$ in Eq.~(\ref{eq:VoscIntPic}) commutes at different times, i.e., $[ \hat{V}_{\rm osc}(t), \hat{V}_{\rm osc}(t') ] = 0$, the general solution of Eq.~(\ref{eq:masterEq}) is given by
\begin{equation}
	\label{eq:mastersol}
	\hat{\rho}(t) = \hat{U}(t,t_0)\hat{\rho}(t_0)\hat{U}^{\dagger}(t,t_0),
\end{equation}
where
\begin{equation}
	\label{eq:timevol}
	\hat{U}(t,t_0) = \exp{\left( -\frac{i}{\hbar}\int_{t_0}^{t}\hat{V}_{\rm osc}(t')dt' \right)}
\end{equation}
is the time evolution operator in the interaction picture.

To numerically solve this equation, we followed the time discretization described in Refs.~\cite{songolo2018nonstandard,songolo2023strang} using the Strang splitting scheme. Since we initially have a BEC in equilibrium, we considered that all population starts in the ground state at $t = 0$. Then, we use Eq.~(\ref{eq:masterEq}) to obtain the evolution of the population (or transition probability) of each state, as well as the coherence between them over time. We can construct resonance curves, as we did with the variational method. The main distinction is that instead of employing the BEC radii, we use the populations of the three modes. This simple model presents a versatile framework, provided that we compute the Rabi-like frequencies associated with the distinct applied excitations. Unlike in the case of atomic states, this model has no decay mechanisms, although their inclusion could possibly lead to a better description in cases with larger excitation amplitudes.

One approach to compute the Rabi-like frequencies is to use Eq.~(\ref{eq:defRabif}), which requires the evaluation of the matrix element
\begin{equation}
	\bra{i}\hat{z}^2\ket{j} = \int_{-\infty}^{+\infty} \psi_i^{*}z^2\psi_jd^3\mathbf{r},
\end{equation}
where $\psi_i$ and $\psi_j$ are the wave functions representing each coherent state in the model. However, defining the required wave functions explicitly is not straightforward.

The method we employed in this work consists of scanning a wide interval of Rabi-like frequencies and finding the values that best reproduce the resonance curves obtained using the variational method, as presented in Fig.~\ref{fig:ress_vari}. To achieve that, we minimized the mean squared error~\cite{draper1998applied} with respect to the Rabi-like frequencies,
\begin{flalign}
	&\chi^2(\{\Omega_{ij}\}) = \frac{1}{\bar{\omega}^{max}_{exc} -\bar{\omega}^{min}_{exc}}\times \nonumber\\
    &\int_{\bar{\omega}^{min}_{exc}}^{\bar{\omega}^{max}_{exc}} \left[ \bar{U}(\bar{\omega}_{exc}) - \tilde{\rho}_{ii}(\{\Omega_{ij}\},\bar{\omega}_{exc}) \right]^2 d\bar{\omega}_{exc},
	\label{eq:error}
\end{flalign}
where $\bar{U} = \bar{R}_r/max(\bar{R}_r)$ and $\tilde{\rho}_{ii} = \rho_{ii}/max(\rho_{ii})$ are normalized such that the amplitude of the main peak corresponds to the unit.

\begin{figure}[!htb]
	\centering
	\includegraphics[scale = 0.3]{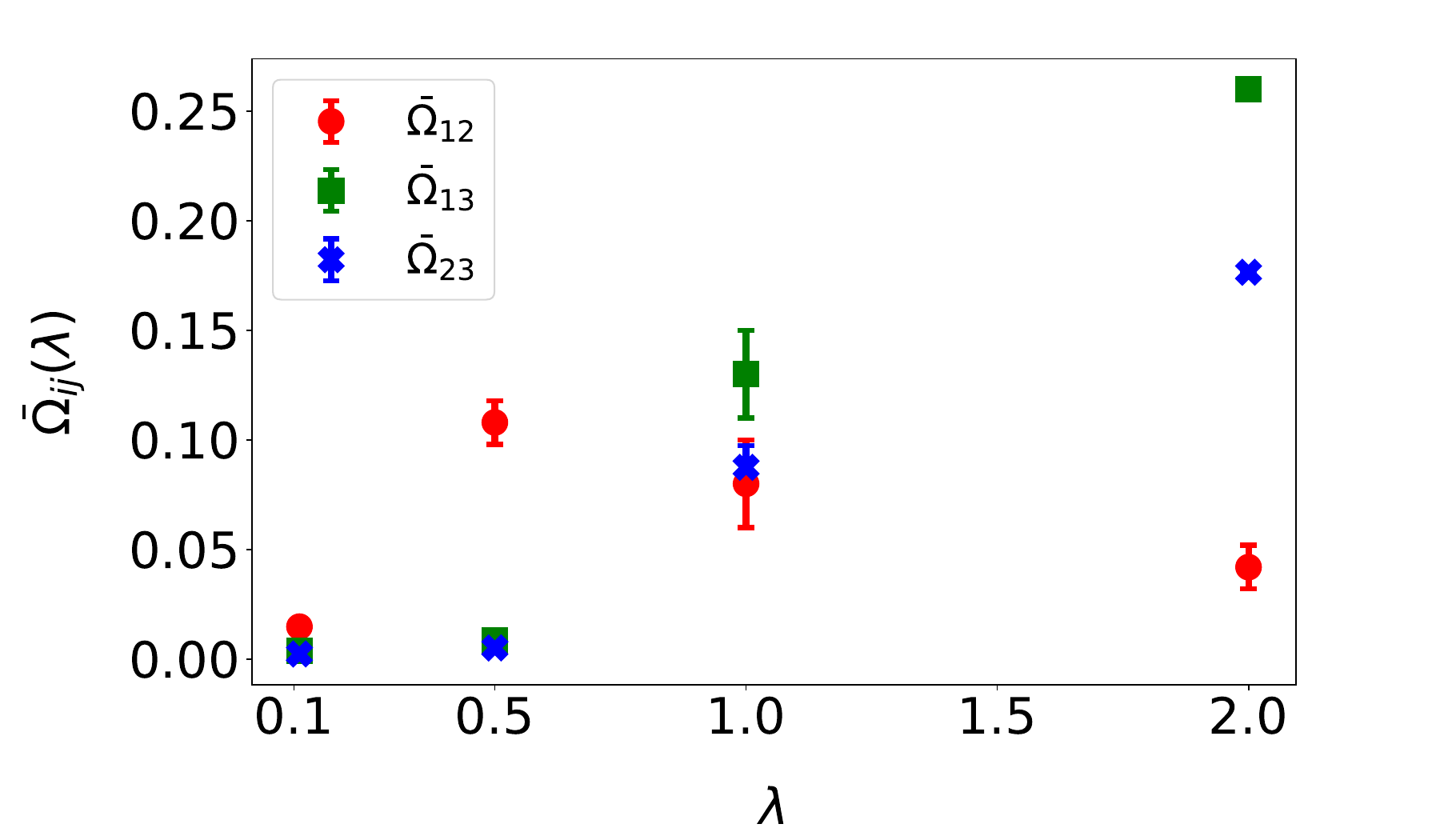}
	\caption{Anisotropy factor $\lambda$ dependence of the Rabi-like frequencies corresponding to the data of Tab.~\ref{tab:tabRabi}.}
	\label{fig:rabifreq}
\end{figure}

We report the resulting Rabi-like frequencies in Tab.~\ref{tab:tabRabi} and plot them as a function of the anisotropy in Fig.~\ref{fig:rabifreq}, where it is possible to see that they do not have a trivial dependence on $\lambda$. For $\lambda = 1/9$, all the Rabi frequencies exhibit relatively low values, confirming that the system is barely excited, as expected from the variational results. The Rabi quadrupole transition frequency $\widebar{\Omega}_{12}$ is significantly larger than the other frequencies for values of $\lambda < 1$, which agrees with the results obtained using the variational approach, where the quadrupole mode is predominantly excited. However, for $\lambda=1$, all the Rabi-like frequencies have similar magnitudes. Thus, the breathing and the quadrupole modes are equally likely to be populated. Indeed, this agrees with the result obtained through the variational method, where the breathing mode played a significant role in the system, unlike the case for $\lambda=1/9$ and $\lambda=1/2$. Notably, $\bar{\Omega}_{12}$ is larger than the one for $\lambda=1/9$ and similar to the one for $\lambda=1/2$. This indicates that the excitation of the quadrupole mode for $\lambda=1$ and $1/2$ will be similar but greater than that in the case of $\lambda=1/9$, also in agreement with the resonance curves of the variational method along the radial direction.

\begin{table}[!htb]
	\caption{Rabi-like frequencies of the three transitions contributing to the population of the quadrupole ($\widebar{\Omega}_{12}$) and breathing ($\widebar{\Omega}_{13}$ and $\widebar{\Omega}_{23}$) modes obtained from the three-level model for different values of the anisotropy factor $\lambda$. We employed the same excitation parameters as the ones used with the variational method.}
	\begin{center}
		\renewcommand{\arraystretch}{1.3}
		\begin{tabular}{lllll}
			\hline
			$\lambda$  & $\widebar{\Omega}_{12}$ & $\widebar{\Omega}_{13}$ & $\widebar{\Omega}_{23}$  \\ \hline
			$1/9$      & $ 0.0149(1)$  & $0.004(1)$   & $0.002(1)$  \\ %\hline
			$1/2$      & $0.11(1)$     & $0.009(1)$   & $0.0052(2)$  \\ %\hline
			$1$        & $0.08(2)$      & $0.13(2)$     & $0.09(1)$   \\ %\hline
			$2$        & $0.04(1)$     & $0.260(1)$   & $0.177(1)$   \\ \hline
		\end{tabular}
	\end{center}
	\label{tab:tabRabi}
\end{table}

\begin{figure*}[t]
	\centering
	\includegraphics[width=400pt]{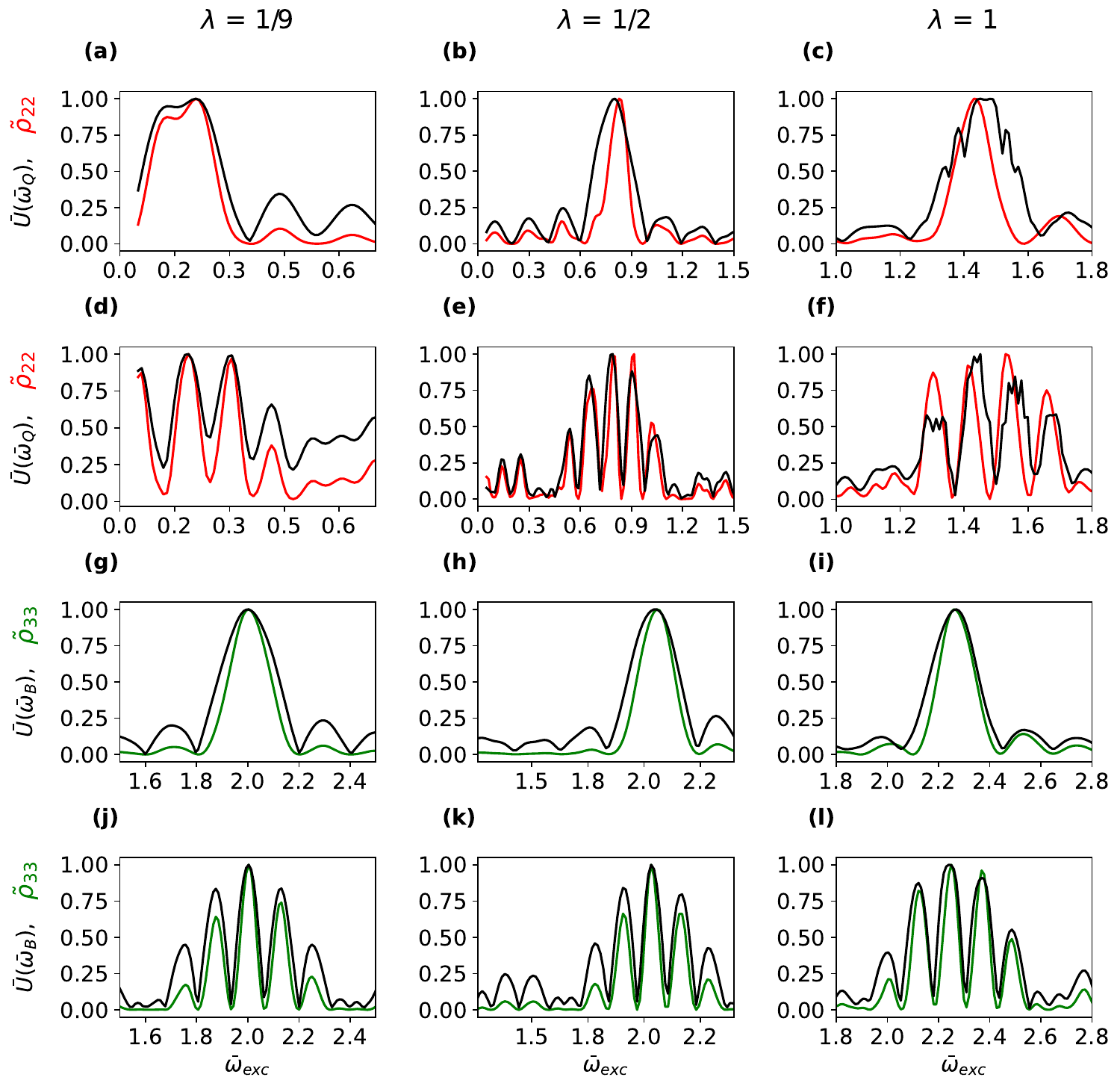}
	\caption{Comparison between the normalized resonance curves of the quadrupole (red) and breathing (green) modes obtained using the three-level model and those employing the variational method (black). The Rabi-like excitation is presented in panels (a)-(c) and (g)-(i), and the Ramsey-like protocol in (d)-(f) and (j)-(l). Each column corresponds to a different trap anisotropy, $\lambda=1/9$, $1/2$, and 1.}
	\label{fig:ress_vari3lvl}
\end{figure*}

We present the resulting resonance curves in Fig.~\ref{fig:ress_vari3lvl}, where we compare the curves obtained with the three-level and the variational methods, adopting the same excitation parameters ($\widebar{A} = 0.05$ and $\tau_{\rm exc} = \tau_{\rm free} = 32$). Notably, the three-level model can qualitatively reproduce the same patterns we obtained using the variational model. The frequencies of the collective modes are also in agreement with the prediction of Eq.~(\ref{eq:wq}) and the ones extracted from the variational resonance curves, as can be seen in Tab.~\ref{tab:tabw}.

\begin{figure}[!htb]
	\centering
	\includegraphics[scale = 0.6]{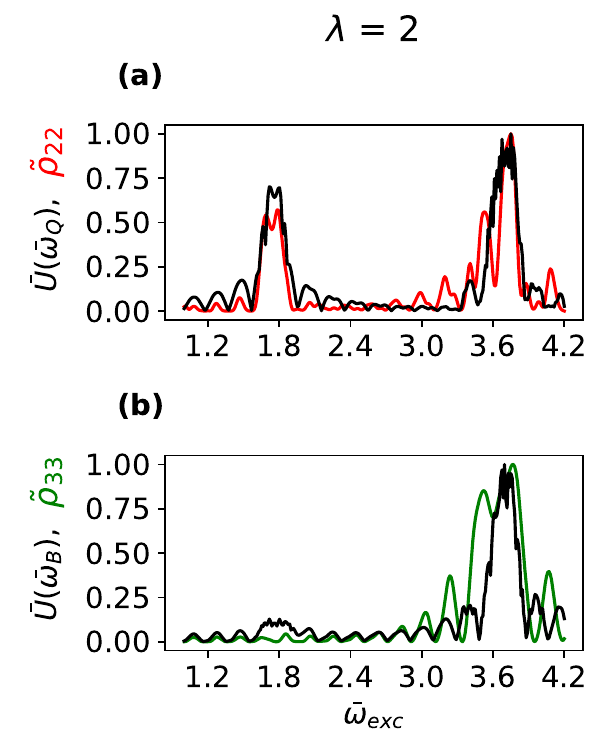}
	\caption{Comparison between the normalized resonance curves of the quadrupole (red line) and breathing (green line) modes obtained from the three-level model and those obtained from the variational method (black line) considering the Rabi excitation protocol for $\lambda=2$.}
	\label{fig:vari3lvl_lamb2}
\end{figure}

We should highlight some important differences between the variational and three-level models, which are even more evident for $\lambda \geq 1$ values. While the three-level model depicts a linear combination of coherent states by construction, the variational model captures additional nonlinear modes as long as they adhere to the Thomas-Fermi profile, Eq.~(\ref{eq:trial}).
Consequently, the variational approach captures nonlinear interactions beyond the superposition of collective modes, leading to spectral broadening in the resonance curves. This broadening arises due to energy redistribution among different collective modes through nonlinear coupling. In contrast, the three-level model assumes a discrete set of non-interacting modes and does not incorporate these nonlinear effects. Nonlinear mode coupling becomes more significant for high-energy excitations, leading to deviations from the idealized few-mode description. This behavior is observed in Figs.~\ref{fig:ress_vari3lvl}(c) and (f) for both Rabi and Ramsey excitations with $\lambda = 1$, and in Figs.~\ref{fig:vari3lvl_lamb2}(a) and (b) for the Rabi excitation with $\lambda=2$.

When the system leaves the linear regime, the interference pattern is no longer well-defined, which happens for $\lambda=2$ with the Ramsey-like excitation. In analogy with an atom being irradiated by a laser beam, the system may undergo the so-called Landau-Zener tunnelling~\cite{shevchenko2010landau,wittig2005landau} at high Rabi frequencies, in which the energy of the system is exchanged between the available levels through a non-adiabatic process. This may result in a breakdown of the coherent dynamics of the system, possibly explaining what we have observed in this context. However, as the system maintains a degree of coherence using the Rabi method for $\lambda=2$, it was still possible to estimate a set of Rabi frequencies, which makes the resonance of the three-level system roughly match the variational one. In this case, the Rabi frequencies are at least ten times larger than those for $\lambda=1$. Furthermore, the width calculated from the resonance curve using the three-level model, particularly when fluctuations play a significant role [Figs. \ref{fig:ress_vari3lvl}(c) and (f), and \ref{fig:vari3lvl_lamb2}(a) and (b)], exhibits lower values compared to those obtained via the variational method (Tab.~\ref{tab:tabw}), which reinforces that while the variational method accommodates other types of fluctuations, the three-level model remains coherent by construction.

Comparing Figs.~\ref{fig:vari3lvl_lamb2}(a)-(b), an inversion in the magnitude of the quadrupole and breathing modes is evident, if compared to the $\lambda\leq 1$ cases. For $\lambda=2$, the quadrupole mode is even more excited at the frequency resonant with the breathing mode frequency, as seen in Fig.~\ref{fig:vari3lvl_lamb2}$(a)$, indicating its strong coupling to the breathing mode in the oblate geometry. Therefore, we conclude that the Rabi frequency strongly depends on $\lambda$ since changing the trap geometry will significantly impact the coupling between modes.

\section{Coherent population control of collective modes}
\label{sec:coherent_control}

The framework provided by the three-level model allows us to approach the dynamics of the collective modes under the light of coherent control, which has been widely used in spectroscopy experiments~\cite{zanon2018composite,silberberg2009quantum}. The idea is to use the framework set forth by Rabi and Ramsey to excite and study the collective modes individually.

The three-level model contains three possible transitions. Thus, we can select a specific transition of interest and tune the external frequency to resonate with it. Moreover, given our understanding of the Rabi frequencies, we can strategically apply either a $\pi$-- or $\pi/2$--pulse to manage the population dynamics of the desired coherent mode as has been done in several interferometry experiments~\cite{sterr1992magnesium, bertet2001complementarity,wan2016free}. A $\pi$--pulse consists in the application of a resonant pulse with a duration of $t_{\rm exc} = \pi/\Omega_{ij}$ resulting in a complete transfer of population from one quantum state $i$ to another one $j$. Similarly, a $\pi/2$--pulse is a resonant pulse with a duration of $t_{\rm exc}=\pi/(2\Omega_{ij})$ that transfers half of the population and places the system into a superposition state. When two $\pi/2$--pulses are applied sequentially with a time interval in between, the system accumulates a phase difference, which can lead to interference effects.

\begin{figure*}[!htb]
	\centering
	\includegraphics[width=350pt]{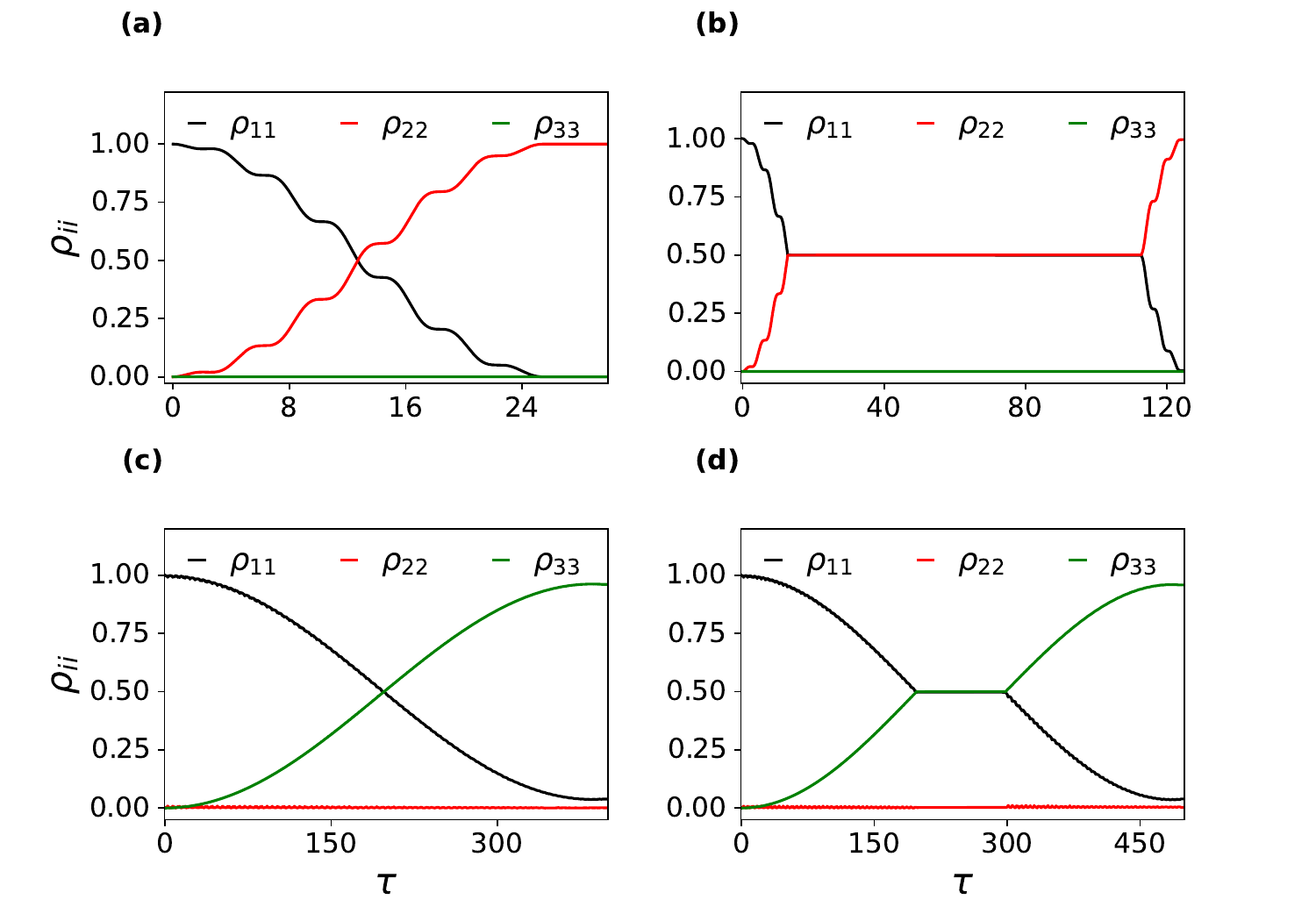}
	\caption{Coherent control of the quadrupole $(a-b)$ and breathing $(c-d)$ mode populations for $\lambda = 1/2$ through the application of a $\pi$-pulse [$(a)$ and $(c)$] and two $\pi/2$-pulses [$(b)$ and $(d)$]. The quantities $\rho_{11}$, $\rho_{22}$, and $\rho_{33}$ represent the populations of the ground state, quadrupole mode, and breathing mode, respectively.}
	\label{fig:coherentcontrol05}
\end{figure*}

Figure~\ref{fig:coherentcontrol05} illustrates a successful attempt at population control of the quadrupole and breathing modes for $\lambda = 1/2$ using both Rabi and Ramsey excitation protocols. To exclusively excite the quadrupole mode ($\rho_{22}$), we employed the Rabi protocol [Fig.~\ref{fig:coherentcontrol05}(a)], where we set the external frequency to be resonant with the quadrupole frequency. The excitation was applied with an amplitude $\bar{A} = 0.05$ over a duration $\tau_{\rm exc}=\pi/\bar{\Omega}_{12}$, corresponding to a $\pi-$pulse. In the Ramsey scheme [Fig.~\ref{fig:coherentcontrol05}(b)], we maintained the same excitation amplitude and we applied two $\pi/2-$pulses, each lasting $\tau_{\rm exc} = \pi/2\bar{\Omega}_{12}$, separated by a free evolution time $\tau_{\rm free} = 100$. Using both protocols, we successfully transferred the population from the ground state to the quadrupole mode. Similarly, Figs.~\ref{fig:coherentcontrol05}(c) and (d) illustrate the same process for the breathing mode, with the external frequency adjusted to match its frequency, and the duration of pulses was determined by the Rabi-like frequency $\bar{\Omega}_{13}$.

A rapid frequency component contributes to this dynamics, particularly in the quadrupole case. This contribution comes from the $e^{i(\omega_{\rm exc} + \omega_{Q,B})t}$ term, which arises when we consider the time evolution of the density operator $\hat{\rho}$~\cite{nielsen2010quantum}. In this case, since the rotating wave approximation~\cite{breuer2002theory,nielsen2010quantum} (RWA) was not assumed in the numerical calculations, we end up with this behavior. Typically, the RWA is valid when this term oscillates much faster than the Rabi oscillations, then averaging to zero, as observed in the coherent control of the breathing mode, when each period of the Rabi oscillations takes a considerable time. However, this term is not much faster than the Rabi oscillations for the quadrupole mode. When this effect is more pronounced, coherent control can be affected; nonetheless, this did not impact the results for $\lambda=1/2$.

\begin{figure*}[!htb]
	\centering
	\includegraphics[width=300pt]{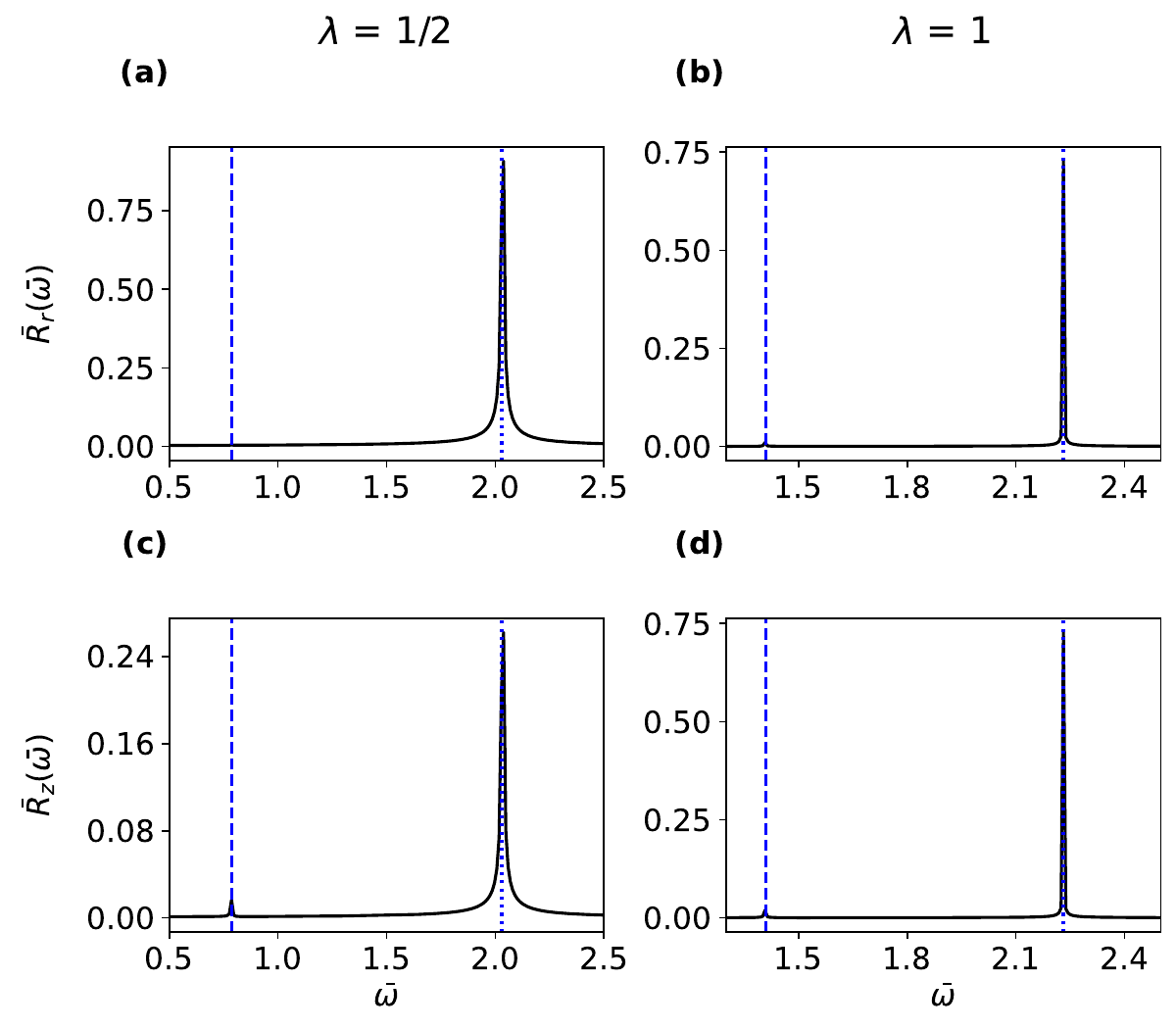}
	\caption{Fourier transforms of the BEC radii oscillations obtained from the variational method for $\lambda=1/2$ $(a-b)$ and for $\lambda=1$ $(c-d)$ when a Rabi pulse with a duration of $\tau_{\rm exc} = \pi/\widebar{\Omega}_{13}$ is applied. The Rabi-like frequency $\widebar{\Omega}_{13}$ was obtained using the three-level model. The blue dashed and dotted lines correspond to the expected frequency for the quadrupole and breathing modes, respectively, according to Eq.~(\ref{eq:wq}).}
	\label{fig:Varipipulse}
\end{figure*}

To illustrate and validate the success of this model in capturing the population dynamics of the collective modes, we can apply a $\pi$-pulse using the variational method with the Rabi-like frequencies obtained from the three-level model. The excitation takes place along the $z$-axis, aligned with the elongated direction of the BEC, favoring the excitation of the quadrupole mode for configurations with $\lambda < 1$. However, even under these circumstances, when the external frequency is set to be resonant with the breathing transition and a $\pi$--pulse of duration $\tau_{\rm exc} = \pi/\widebar{\Omega}_{13}$ is applied, the result is the excitation of the breathing mode. Figure~\ref{fig:Varipipulse} presents the Fourier transform of the oscillations in the BEC radii when a $\pi$-pulse is applied, for $\lambda = 1/2$ and $\lambda = 1$. In contrast to the scenarios presented in Fig.~\ref{fig:FFT_osc} for $\lambda=1/2$ and $1$, we observe the breathing mode being predominantly excited, showing the successful manipulation of the population in each state.

This outcome suggests the feasibility of coherent control within this system, offering valuable insights for experimentalists investigating specific collective modes. Moreover, our results highlight the well-known fact that collective modes in a zero-temperature BEC exhibit coherence, allowing their dynamics to be described using techniques from quantum optics. Exploring this connection could be relevant for future studies on coherent control of macroscopic quantum states in interacting many-body systems.

Considering experimental factors such as inherent line broadening in the external excitation, the quadrupole mode might be excited even under conditions intended to excite the breathing mode exclusively. To ensure that most of the population resides in the breathing mode, an additional $\pi$-pulse should be implemented, specifically targeting the transition from the quadrupole to the breathing mode. This process mirrors protocols seen in atomic physics experiments known as repump protocols \cite{long2018magnetic}.

\begin{figure}[!htb]
	\centering
	\includegraphics[scale = 0.5]{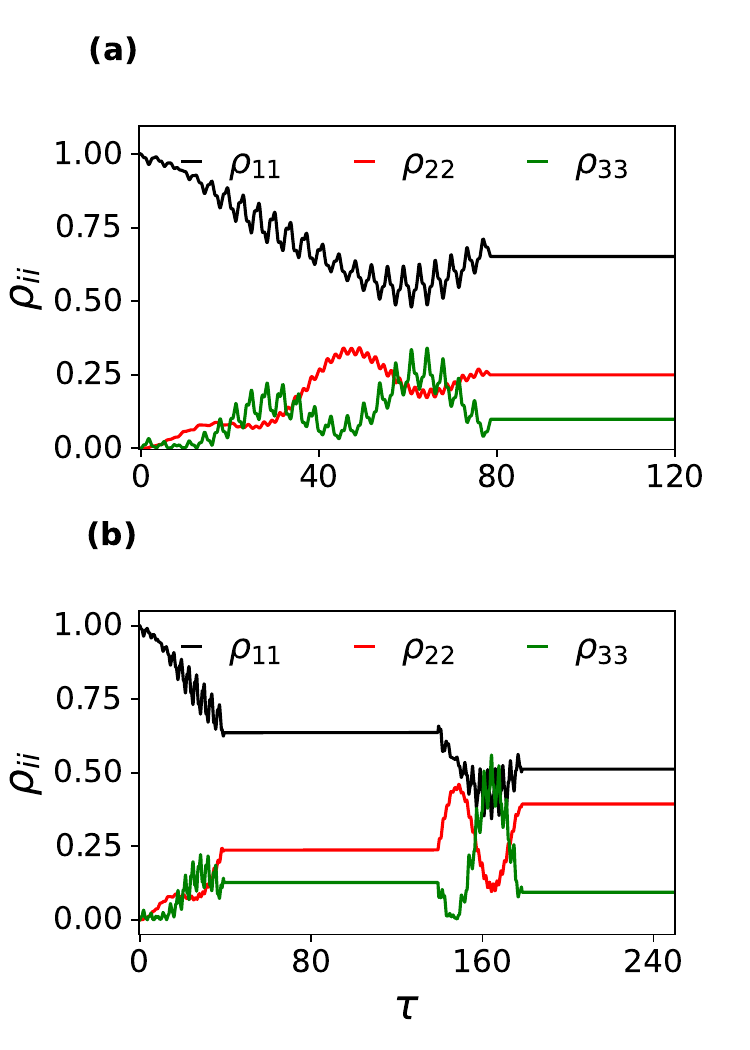}
	\caption{A failed attempt to coherently control the quadrupole population by applying a $\pi$-pulse $(a)$ and two $\pi/2-$pulses $(b)$ for $\lambda=2$. Since the perturbation is relatively large, the influence of nonlinear} modes prevents a coherent manipulation of the modes.
	\label{fig:quadcontrol2}
\end{figure}

Although it is expected, we should point out that coherent control fails for arbitrarily large excitation amplitudes. For higher values of $\lambda$, the influence of nonlinear modes is significant, as discussed in Sec.~\ref{sec:three_level}, and coherent control of the modes is not possible. For $\lambda = 2$, this can be seen in Fig.~\ref{fig:quadcontrol2}, where the attempt of total transfer to the quadrupole mode failed. The Rabi frequencies associated with $\lambda=2$ are relatively high (see Tab.~\ref{tab:tabRabi}), of the order of the external frequency, and the fast rotation term dominates, thus making the coherent control impossible for this case.

\section{Conclusions}
\label{sec:conclusions}

In this work, we explored the coherence properties of Bose-Einstein condensates by employing external pulses inspired by the Rabi and Ramsey protocols to analyze the collective modes of the system. We showed that collective modes can exhibit Ramsey fringes in their resonance curves, allowing for more precise measurements of their frequencies when coherence is maintained. This approach opens up the possibility of applying these excitation protocols to control the population of collective modes in experiments more precisely.

We also proposed an alternative way of describing the dynamics of collective modes as a three-level system, achieving consistent results with those obtained using the variational method. Given the similarities between BECs and coherent radiation fields~\cite{andrews1997observation, mewes1997output, bloch1999atom, jos1, jos2, tavares2017matter, madeira2021cold, modugno2010anderson}, the Ramsey protocol holds promise for manipulating quantum states through interference phenomena.

These findings offer an alternative perspective for observing collective mode dynamics and present a new framework for describing quantum collective modes of a many-body system with simpler approaches that capture the relevant physics. In future works, we intend to explore further the transition from \textcolor{blue}{linear} to \textcolor{blue}{nonlinear} regimes, which can provide a deeper understanding of the dynamics of a quantum system moving away from equilibrium and possibly reaching disordered states.

\section{Acknowledgments}
The authors thank H.~Perrin and R.~Dubessy for interesting discussions and valuable insights. This work was supported by the São Paulo Research Foundation (FAPESP) under the grants 2013/07276-1, 2014/50857-8, 2023/04451-9, and 2024/04637-8. L.A.M. acknowledges the support from Coordenação de Aperfeiçoamento de Pessoal de Nível Superior - Brasil (CAPES) - Finance Code 88887.684421/2022-00, 88887.822682/2023-00, and 88887.999663/2024-00. Texas A\&M University is acknowledged.

\section*{Data Availability Statement}

The data that supports the findings of this study are openly available in the Zenodo repository at \url{http://doi.org/10.5281/zenodo.14206482}, reference number~\cite{alvares_machado_2024_14206482}.

\bibliography{references}

\end{document}